\documentclass[francais]{gretsi}
\usepackage{url}

\usepackage[utf8]{inputenc}

\usepackage[T1]{fontenc}

\usepackage{amsmath, amssymb, amsfonts}

\usepackage{siunitx}
\DeclareSIUnit \percent {\, \%}

\usepackage{multirow}

\newcommand{\bvv}{\vspace{-.05in}}

\RequirePackage{amsmath}
\RequirePackage{xspace}
\def\sbv#1{\ensuremath{\mathbf{#1}}}               
\def\XS{\xspace}
\def\piv  {{\sbv{\pi}}\XS} 

\def\Av{{\sbv{A}}\XS}

  \def\ev{{\sbv{e}}\XS}
  
\def\Gv{{\sbv{G}}\XS}  
\def\Hv{{\sbv{H}}\XS}

\def\Lv{{\sbv{L}}\XS}

  \def\pv{{\sbv{p}}\XS}

  \def\sv{{\sbv{s}}\XS}

  \def\zv{{\sbv{z}}\XS}

\begin{document}


\titre{Analyse comparative d'algorithmes de restauration en architecture dépliée pour des signaux chromatographiques parcimonieux}

\auteurs{
  \auteur{Mouna}{Gharbi}{mouna.gharbi@unigraz.at}{1}
  \auteur{Silvia }{Villa}{villa@dima.unige.it}{2}
  \auteur{Emilie }{Chouzenoux}{emilie.chouzenoux@centralesupelec.fr}{3}
  \auteur{Jean-Christophe}{Pesquet}{jean-christophe.pesquet@centralesupelec.fr}{3}
  \auteur{Laurent}{Duval}{laurent.duval@ifpen.fr}{4}
}

\affils{
  \affil{1}{University of Graz, Department of Mathematics and Scientific Computing, Graz, Austria  }
  \affil{2}{MaLGa, DIMA, Universit\`a degli Studi di Genova, Via Dodecaneso 35, 16146 Genova, Italy}
  \affil{3}{CVN, CentraleSup\'elec, Inria Saclay, University Paris Saclay, France}
  \affil{4}{IFP Energies nouvelles, Rueil-Malmaison, France}
}

\resume{La restauration de données à partir d'observations dégradées,  sous hypothèse de parcimonie, est un champ d'étude  actif. Aux traditionnelles méthodes itératives d'optimisation  s'ajoutent aujourd'hui l'usage de techniques d'apprentissage profond. Le développement des méthodes dépliées bénéficie des avantages  de ces deux familles. Nous réalisons une étude comparative de trois architectures sur des bases de données paramétrées de signaux en chromatographie, mettant en évidence la performance de ces approches, notamment en employant des métriques adaptées à la caractérisation de signaux de pics physico-chimiques.}

\abstract{Data restoration  from degraded observations,  of sparsity hypotheses, is an active field of study. Traditional iterative optimization methods are now complemented by deep learning techniques. The development of unfolded methods benefits from  both families. We carry out a comparative study of three architectures on parameterized chromatographic signal databases, highlighting the performance of these approaches, especially when employing metrics adapted to physico-chemical peak signal characterization.}

\maketitle


\section{Introduction}
\begin{figure}[htb]
	\centering
		\includegraphics[height=2.8cm,keepaspectratio]{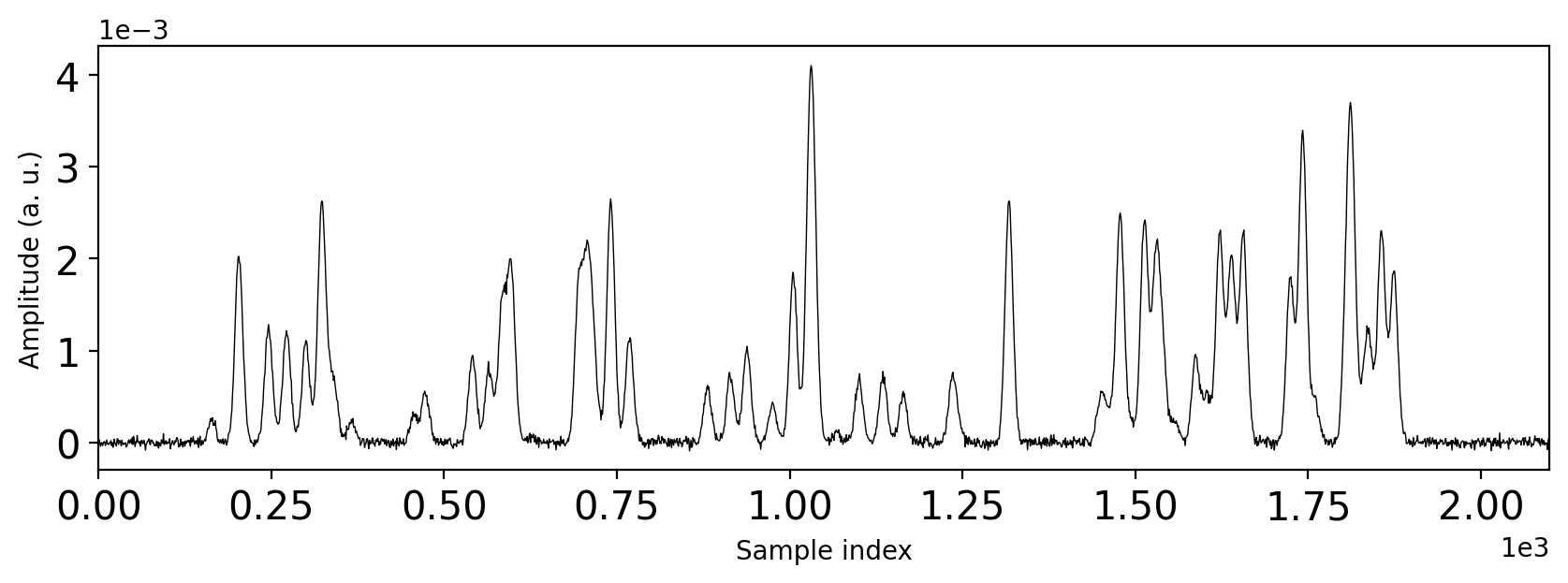}	
	\caption{Observation bruitée $\zv$ d'une somme  de pics  $\pv$.}
	\label{fig:GR-DEG}
\end{figure}

\begin{figure}[htb]
	\centering
	\includegraphics[height=2.8cm,keepaspectratio]{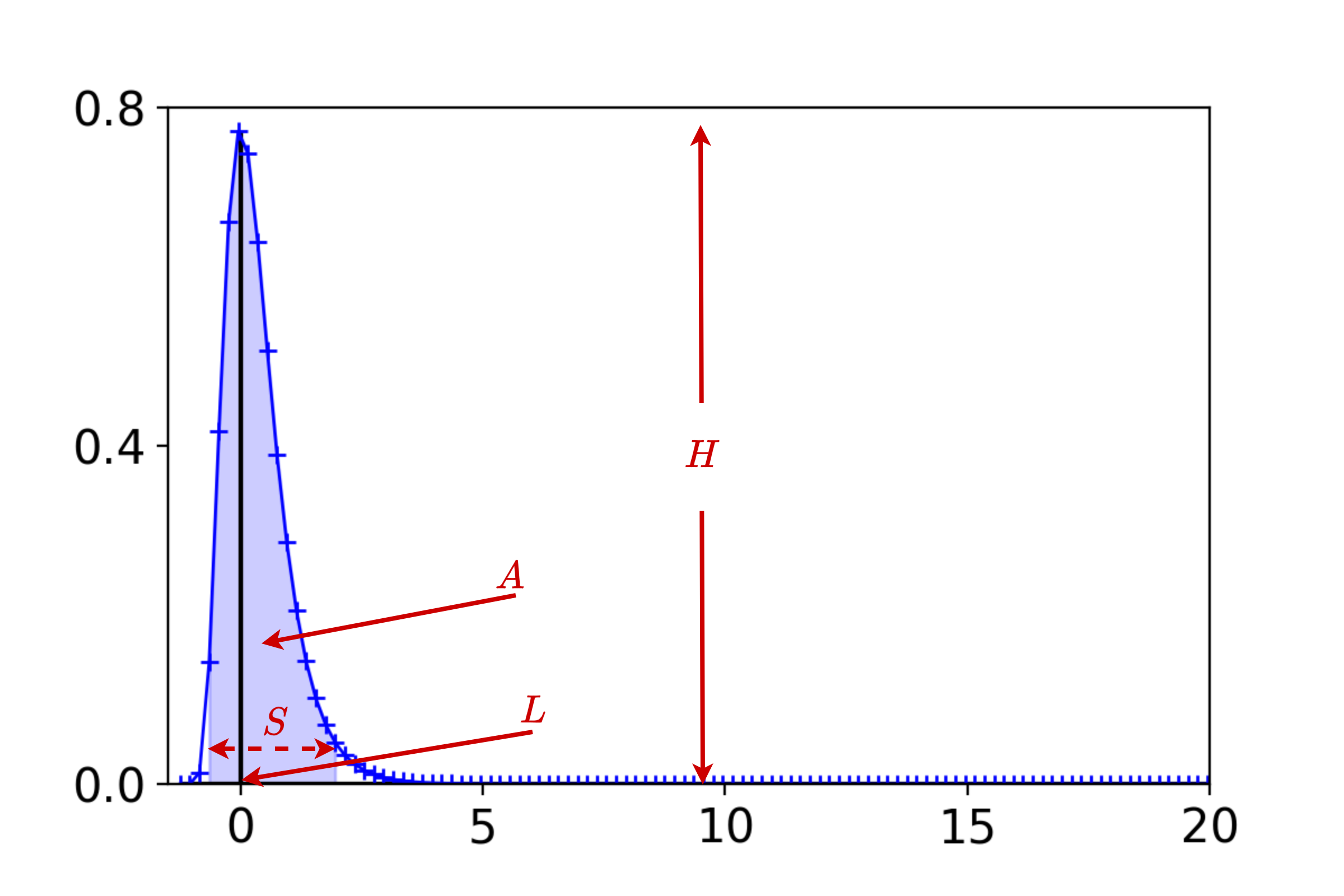}
	\caption{Caractérisation d'un pic : hauteur ($H$),  aire ($A$),  localisation ($L$).  Support ($S$) et aire colorée (bleu)  déterminés  en fonction d'un seuil sur $H$  (ici $H/20$).}
	\label{fig:chem_char}
\end{figure}

Ce travail\footnote{Projets supports : EU H2020-MSCA-ETN TraDE-OPT-861137; ERC Starting Grant MAJORIS ERC-2019-STG850925; AFOSR  FA8655-22-1-7034 ; ERC 
	Consolidator Grant SLING 819789; EU H2020-MSCA-RISE
	NoMADS-777826.} s'intéresse à une classe de signaux parcimonieux communément décrits comme formés d'une somme de pics relativement étroits ($\pv$), potentiellement dégradés par un noyau de lissage et du bruit ($\ev$), résultant en une observation $\zv$. Ce modèle est courant en chimie analytique, notamment pour le traitement de données de  type chromatographique,  représentées dans la figure \ref{fig:GR-DEG}. On considère connue une forme paramétrique  $\piv$ de pic, illustré en figure 	\ref{fig:chem_char}. Le problème inverse à résoudre est l'estimation d'un train d'impulsions parcimonieux $\sv$ (impulsion représentée par une barre noire verticale, figure \ref{fig:chem_char}). La prise en compte de la réponse globale du système d'acquisition se fait par un opérateur linéaire $\Gv$, par exemple un lissage ou opérateur de flou \cite{Liu2012}, une transformée de Fourier \cite{Delsuc2021} ou de Laplace \cite{Christian2024}. Il en résulte : 

\begin{equation}
	\zv=\Gv \pv+\ev = \Gv (\piv \ast \sv)+\ev \,.
	\label{eq:model}
\end{equation}

La résolution du problème inverse consiste en l'obtention d'une estimation de  $\pv$ (ainsi que de   $\sv$ si possible), connaissant $\zv$ et $\Gv$, avec des hypothèses sur le bruit $\ev$, et bien entendu en promouvant des solutions parcimonieuses. Pour que les résultats restent interprétables et utiles pour les chimistes, il est important de considérer des indicateurs de qualité complémentaires aux quantités usuelles. Dans le modèle paramétrique  considéré, l'estimation des facteurs de forme (notamment :  hauteur ($H$),  aire ($A$) et   localisation ($L$)) pour chaque pic est également importante, car ces derniers permettent d'accéder à des informations sur la nature chimique du mélange (type de composé, concentration de chaque espèce du mélange, etc.). 

On distingue  des approches  (i)  basées sur des modèles, employant des algorithmes d'optimisation itératifs et (ii)  d'apprentissage (profond), basées sur des architectures neuronales. Chacune a des lacunes. Un paradigme récent, dit dépliement ou déroulement profond (\emph{deep unfolding/unrolling}) \cite{Hershey2014}, tente de les combler.
Ce travail prolonge  \cite{Gharbi_M_2024_p-mlsp_unrolled_dnssrac}, et s'intéresse  :

\begin{itemize}
	\item au développement d'un simulateur réaliste de bases de données, aux paramètres ajustables (section \ref{sec_exp}) ;
	\item à l'analyse comparative de trois architectures dépliées basées sur des modèles, pour la résolution de problèmes inverses en chimie analytique et à leur formulation (sections \ref{sec_soa} et
	\ref{sec:proposed method}) ; 
	\item à leur évaluation, incluant des métriques HAL (hauteur-aire-localisation) inspirées de la chimie, plus adaptées à cette classe de signaux parcimonieux (section \ref{sec_res}).
\end{itemize}

Les codes proposés sont disponibles sous forme d'une boîte à outils écrite en Pytorch pour satisfaire à des exigences de reproductibilité. 


\section{Modèle de données et évaluation}
\label{sec_exp}

L'équation \ref{eq:model} nous conduit à simuler un modèle additif basé sur la forme de pic asymétrique  et paramétrique à noyau de  Fraser-Suzuki  \cite[p. 104]{Felinger1998}: $\forall x> m-\frac{\sigma_f}{a} $,
\begin{equation}
	\piv(x)\propto \operatorname{exp}\left(-\frac{1}{2a^2}\operatorname{log}\left(1+a\frac{(x-m)}{\sigma_f}\right)^2\right).
\end{equation}
Des pics gaussiens s'obtiennent quand $a\to 0$. La discrétisation des signaux induit les  paramètres suivants :

\begin{itemize}
	\item signal impulsif $\sv$:  nombre d'échantillons $N=2000$;  nombre d'impulsions distinctes   $P$, exprimé relativement à la taille du signal par  $P/N$ ; distance minimale séparant deux pics $d_{\min}$ ;  intensités des pics  tirées aléatoirement (valeur absolue de la réalisation d'une loi normale de moyenne 0 et de variance 1) ;
	\item  noyau de Frazer-Suzuki $\piv$:  largeur de pic $\sigma_f>0$ ;  coefficient d'asymétrie  $a>0$ ;
	\item sources d'altération : 
	bruit additif gaussien de moyenne nulle et d'écart-type $\sigma_{\ev}>0$;
	opérateur de flou gaussien de largeur $\sigma_{\Gv}=1$ (fixé pour cette étude).
\end{itemize}
Pour tester l'influence des paramètres $(N,P,d_{\min},\sigma_f,a,\sigma_{\ev})$, nous proposons sept bases de données (table~\ref{tab:datasets}) dont la complexité relative est résumée en table 	\ref{tab:datasets-doe}.

\begin{table*}[hbt]
	\centering
	\caption{Bases de données: parcimonie relative ($P/N$), distance minimale entre impulsions ($d_{\min}$), largeur de $\pi$ ($\sigma_f$) et facteur d'asymétrie ($a$),  écart-type du bruit ($\sigma_\ev$).}
	\begin{tabular}{|c|c|c|c|||c|c|||c|c|c|c}
		\hline
		Param. &D0&D1&D2&D3&D4&D5 &D6\\
		\hline
		$P/N$ &$1.5 \si{\percent}$& $3\si{\percent}$ & $4.5\si{\percent}$&$1.5\si{\percent}$&$1.5\si{\percent}$& $3\si{\percent}$& $3\si{\percent}$\\
		\hline
		$d_{\min}$&5&$3$&$1$&$5$&$5$&$3$&$3$\\
		\hline
		$\sigma_f$&$0.5$&$0.5$&$0.5$&$0.5$&$0.5$&$0.5$&$0.5$\\
		\hline
		$a$&$0.2$&$0.2$&$0.2$&$0.4$&$0.6$&$0.2$&$0.2$\\
		\hline
		$\sigma_{\ev}$ &$0.02$&$0.02$&$0.02$&$0.02$&$0.02$ &$0.04$&$0.06$\\
		\hline
	\end{tabular}
	\label{tab:datasets}
\end{table*}

\begin{table*}[hbt]
	\centering
	\caption{Variations paramétriques (cf. table \ref{tab:datasets}) sur la parcimonie, l'asymétrie et le bruit, avec trois niveaux de difficulté.}
	\begin{tabular}{|l|l|l|l|}
		\hline
		Difficulté &Faible&Moyenne&Haute\\
		\hline
		Parcimonie $(P/N,d_{\min})$ &D0 $(1.5\si{\percent},5)$& D1 $ (3\si{\percent},3)$ & D2 $(4.5\si{\percent},1)$\\
		\hline
		Asymétrie $(a)$&D0 $(0.2)$&D3 $(0.4)$&D4 $(0.6)$\\
		\hline
		Bruit $(\sigma_{\ev})$ &D1 $(0.02)$&D5 $(0.04)$&D6 $(0.06)$\\
		\hline
	\end{tabular}
	\label{tab:datasets-doe}
\end{table*}

Notre objectif est ensuite de comparer la performance des différents algorithmes. Cependant, il est connu que les métriques "quadratiques" standard se prêtent mal aux signaux parcimonieux, comportant une information entre-pics "inutile". Nous allons supposer que chaque pic possède (dans la vérité-terrain) un support effectif $\mathcal{S}_j$ connu, sur lequel ses caractéristiques HAL (hauteur-aire-localisation) peuvent être évaluées et comparées.
\'Etant donné le signal original $\sv\ast \pi=\pv \in \mathbb{R}^N$ et sa restauration $\widehat{\pv}\in \mathbb{R}^N$, nous calculons dans un premier temps trois métriques usuelles : erreur quadratique moyenne (MSE), rapport signal/bruit (SNR, exprimé en décibels) et sa version tronquée  (TSNR) sur le support. Cette dernière correspond à un rapport signal/bruit évalué uniquement sur l'union  des supports de tous les pics, c'est-à-dire uniquement sur le signal d'intérêt  physico-chimique, $\mathcal{S} = \bigcup\limits_{1 \leq j \leq P} \mathcal{S}_j$:

\begin{align} 
	\text{MSE}(\pv,\widehat{\pv}) &=\frac{1}{n}\| \pv - \widehat{\pv}\|^2,\\
	\text{SNR}(\pv,\widehat{\pv})&= 20 \log_{10}\left(\frac{\|\pv\|}{\| \pv - \widehat{\pv}\|}\right),\\
	\text{TSNR}(\pv,\widehat{\pv})&= 20 \log_{10}\left(\frac{\sum_i|\pv^{(i)}|^2}{\sum_i| \pv^{(i)} - \widehat{\pv^{(i)}}|^2}\right),  \forall i\in  \mathcal{S}.
\end{align}
Si l'introduction du TSNR améliore l'évaluation des performances pour des signaux parcimonieux, cette mesure est  complétée par des indicateurs de morphologie des pics, utiles en chimie.
Nous employons trois quantités pour chaque pic : hauteur ($H$), 
aire sous le pic ($A$),  localisation ($L$), à l'aide de l'information sur le  support essentiel $\mathcal{S}_j$. Elles sont illustrées sur  leur figure~\ref{fig:chem_char}. Les grandeurs  $H$ et  $A$ peuvent être
reliées à la prévalence d'une entité chimique donnée, tandis que  $L$ peut servir à la caractérisation ou la détection. Les valeurs vraies issues du modèle de pic  $\overline{H_{j}}$ et $\overline{L_{j}}$
sont obtenues directement pour chaque pic convolué. Nous faisons l'hypothèse que le support essentiel  $\mathcal{S}_j$ est défini par les indices auxquels l'amplitude du signal est supérieure à une certaine fraction  ($0<\vartheta < 1$) de la hauteur $\overline{H_{j}}$. Ici, on prend $\vartheta = \frac{1}{20}$.
L'aire $\overline{A_{j}}$ est calculée par intégration discrète (méthode des trapèzes) des échantillons supportés par  $\mathcal{S}_j$. En utilisant cette information comme oracle, des quantités analogues sont obtenues sur le signal restauré   $\hat{\pv}$: $\widehat{L_{j}}$, $\widehat{H}_j$ et $\widehat{A_{j}}$.

Ces indicateurs sont résumés dans des vecteurs $\overline{\Hv} = (\overline{H_{j}})_{1 \leq j \leq P}$ et $\widehat{\Hv} = (\widehat{H_{j}})_{1 \leq j \leq P}$ de longueur $P$, en définissant de manière analogue  $\overline{\Av}$,  $\overline{\Lv}$,  $\widehat{\Av}$ et  $\widehat{\Lv}$. Ils permettent d'obtenir des graphes de dispersion comme en figure	\ref{fig:comparison_H_sparsity}. Leur écart plus ou moins prononcé à la linéarité permettant de comparer subjectivement les performances. Cette analyse peut être affinée en distinguant les pics suivant leur taux de recouvrement. En cyan et jaune sont représentés respectivement les pics présentant moins (plus) de 30\% de recouvrement. On voit que les points jaunes sont souvent situés au-dessus de la diagonale. Ceci est probablement lié à l'emprunt d'information entre deux pics adjacents, mal séparés, provoquant une surestimation des hauteurs. Ces observations motivent la définition d'une mesure objective normalisée d'écart des hauteurs :

\begin{equation}
	{\text{NMAE}(\overline{\Hv},\widehat{\Hv})=}   \frac{\sum_{j=1}^P | \overline{H_{j}} - \widehat{H_{j}}|}{\sum_{j=1}^P |\overline{H_{j}}|}.
\end{equation}
Des mesures similaires  sont définies pour l'aire {$\text{NMAE}(\overline{\Av},\widehat{\Av})$} et la localisation {$\text{NMAE}(\overline{\Lv},\widehat{\Lv})$}.


\section{Travaux analogues}
\label{sec_soa}
Nous mentionnons ici quelques travaux en lien avec cette étude, avec des applications en chimie analytique, en commençant par les méthodes d'optimisation itérative. Le débruitage de spectres Raman emploie une régularisation en variation totale  \cite{tripathi2020}. Une approche bayésienne de débruitage et déconvolution d'images spectroscopiques en infrarouge à transformée de Fourier est proposée dans  \cite{nguyen2012}. La reconstruction de données spectroscopiques est obtenue  en incorporant des contraintes de positivité, par régularisation en maximum d'entropie \cite{chouzenoux2010Efficient} ou par méthode de point intérieur primal-dual dans \cite{chouzenoux2013}, plus récemment dans \cite{cherni2020} par méthode  de majoration-minimisation. La restauration de données chromatographiques par moindres carrés alternés est proposée dans  \cite{shi2005}. 
Les méthodes d'apprentissage profond jouent un rôle croissant en chimie, avec des revues génériques \cite{mater2019,chen2020,debus2021}. En chromatographie, notons la	classification de profils d'élution ou la détection automatisée de pics 
\cite{Risum_A_2019_j-talanta_using_dlepcd,Kensert_A_2022_j-chrom-a_convolutional_nnapdrplc}. Le recours à des algorithmes dépliés en chimie n'est pas encore très développé. Nous mentionnons notamment \cite{bobin2019,wang2022,Kervazo_C_2024_p-eusipco_deep_umuabssahu,Gharbi_M_2024_j-sp_unrolled_hqassrs}.


\section{Algorithmes}
\label{sec:proposed method}
Nous considérons trois architectures dépliées, basées sur les algorithmes itératifs : primal-dual proximal, ISTA et semi-quadratique (HQ). Nous rappelons uniquement les problèmes de minimisation, et renvoyons par manque d'espace à  
\cite{Gharbi_M_2024_p-mlsp_unrolled_dnssrac} pour les détails:

\begin{itemize}
	\item  primal-dual déplié (U-PD) :  $\rho>0$ est un hyper-paramètre lié au niveau de bruit. 
	\begin{equation}
		\widehat{x} \in  \arg \min_{x \in \mathbb{R}^n} \|x\|_1\\ \quad \text{s.t.} \quad \|Hx-z\|_2 \leq \rho.
		\label{eq:constrained-opt}
	\end{equation}
	\item  seuillage doux itératif (ISTA) déplié (U-ISTA) :  $\chi$  est un hyper-paramètre couplant les deux termes.
	\begin{equation}
		\label{eq:cost_fun_ISTA}
		\widehat{x} \in  \arg \min_{x \in \mathbb{R}^n} \Big(F(x) = \frac{1}{2}\| H x - z \|_2^2 + \chi \|x\|_1\Big).
	\end{equation}
	\item Semi-quadratique déplié (U-HQ) : 
	\begin{equation}
		\label{eqn:cost_funHQ}
		\widehat{x} \in  \arg \min_{x \in \mathbb{R}^n} \Big(F(x) = \frac{1}{2}\| H x - z \|_2^2 + \Psi(x) \Big),
	\end{equation}
	avec $ \Psi(x) = \sum_{i=1}^n \psi(x_i)$ et 
	\begin{multline}
		\label{eq:hybrid}
		(\forall t \in \mathbb{R}) \quad \psi(t)= \lambda_1 \delta_1  \left(|t|-\delta_1 \log \left(\frac{|t|}{\delta}_1 +1\right)\right) + \\
		\lambda_2\frac{\delta_2^2}{2} \log \left( 1 + \frac{t^2}{\delta_2}\right).
	\end{multline}
\end{itemize}


\section{Résultats}

\label{sec_res}
Nous comparons ces trois algorithmes dépliés pour un nombre de couches fixé à $K=4$. La table  \ref{tab:arb_layers_MSE_SNR_TSNR} rassemble les métriques  MSE, SNR, TSNR et HAL. 
\textbf{U-HQ} présente de manière évidente  des résultats supérieurs à ceux de  \textbf{U-PD} et \textbf{U-ISTA}. Notons que cette performance se répercute en coût de calcul, le temps moyen d'exécution par jeu de données étant respectivement de 0,09, 0,009 et 0,003 secondes.

L'intérêt de la constitution de bases de données paramétrables est d'analyser plus précisément l'impact d'un facteur en particulier. Si l'on regarde la parcimonie (ligne 1, table \ref{tab:datasets-doe}), on observe pour les jeux de données D0, D1 et D2 que l'accroissement du nombre d'impulsions (d'amplitude non nulle) dégrade la qualité de reconstruction. Pour les trois méthodes, la mesure NMAE augmente pour $\Hv$, $\Av$ et $\Lv$. 
Les méthodes dépliées ont des difficultés croissantes à estimer correctement hauteur, amplitude et localisation, à mesure que les signaux sont plus denses en pics (moins parcimonieux). On remarque également que les métriques globales SNR et TSNR ne sont pas des indicateurs fiables pour ce phénomène, car leurs valeurs sont assez similaires pour les trois ensembles de données.

\begin{figure}[h!]
	\centering
	\begin{tabular}{@{}c@{}c@{}c@{}}
		
		\includegraphics[width=1.1in,keepaspectratio]{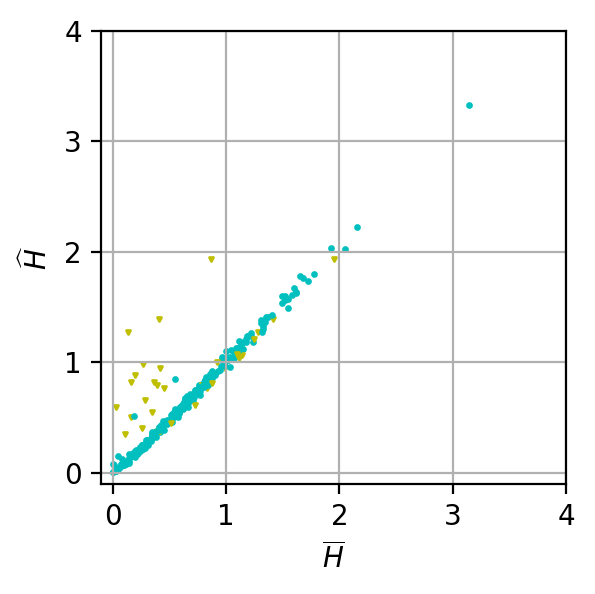}&
		\includegraphics[width=1.1in,keepaspectratio]{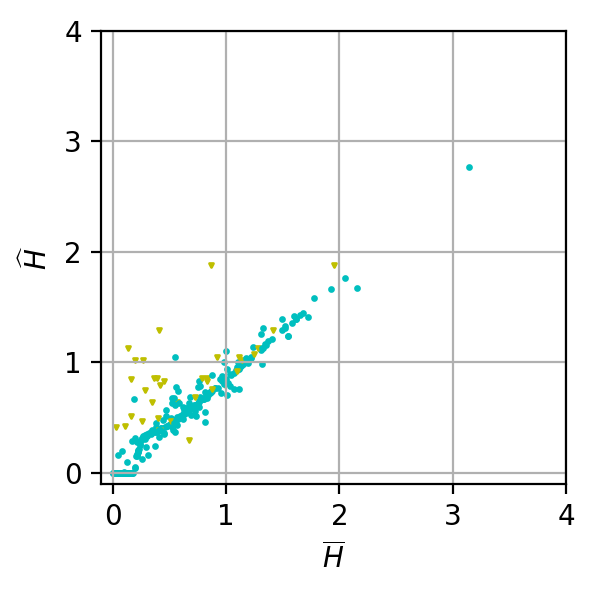}&
		\includegraphics[width=1.1in,keepaspectratio]{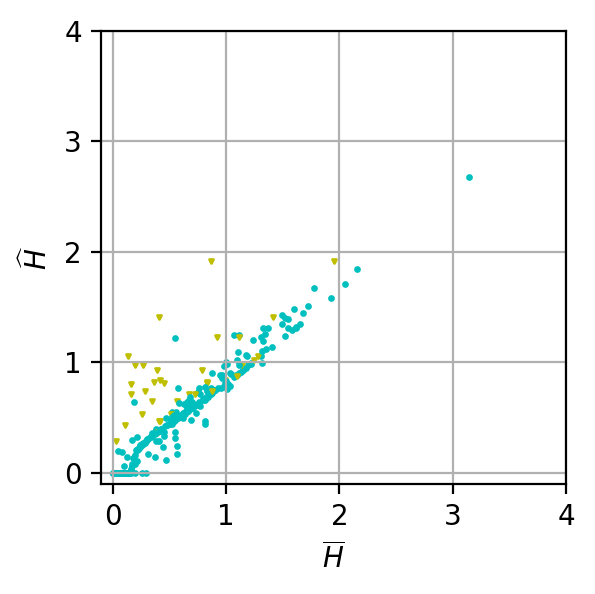}\\
		
		\includegraphics[width=1.1in,keepaspectratio]{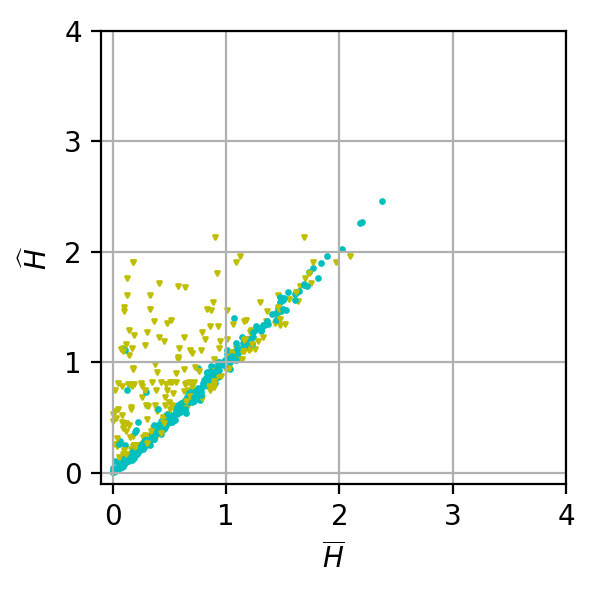}&
		\includegraphics[width=1.1in,keepaspectratio]{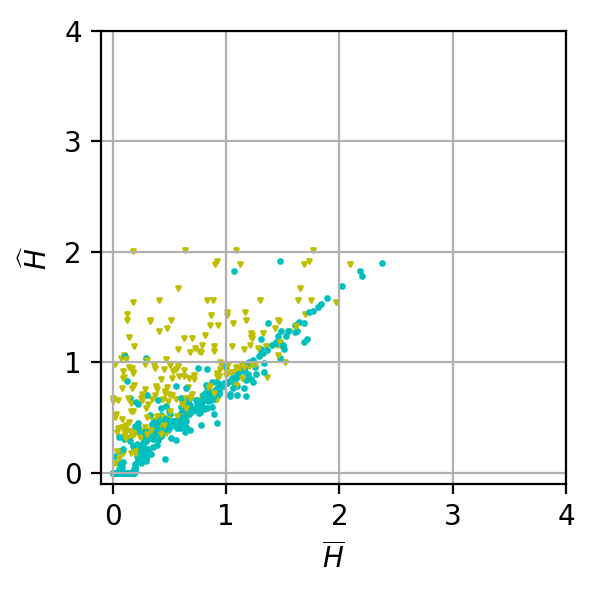}&
		\includegraphics[width=1.1in,keepaspectratio]{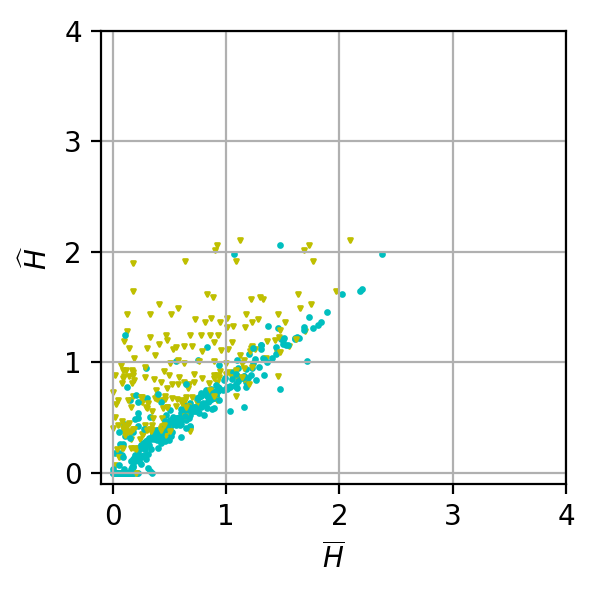}\\
		
		\includegraphics[width=1.1in,keepaspectratio]{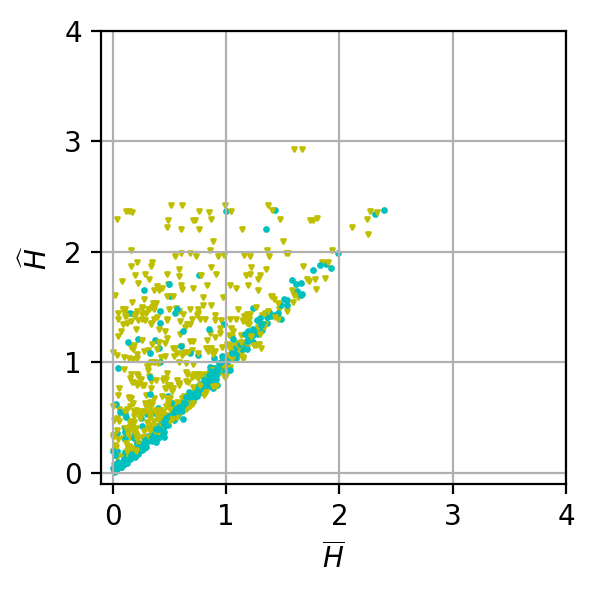}&
		\includegraphics[width=1.1in,keepaspectratio]{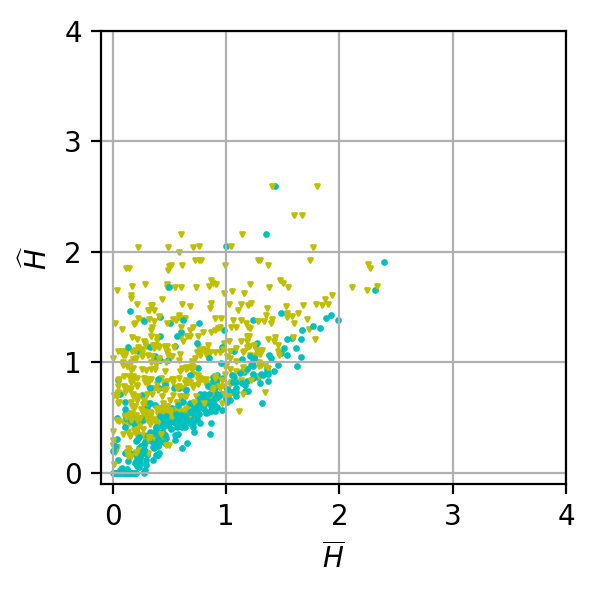}&
		\includegraphics[width=1.1in,keepaspectratio]{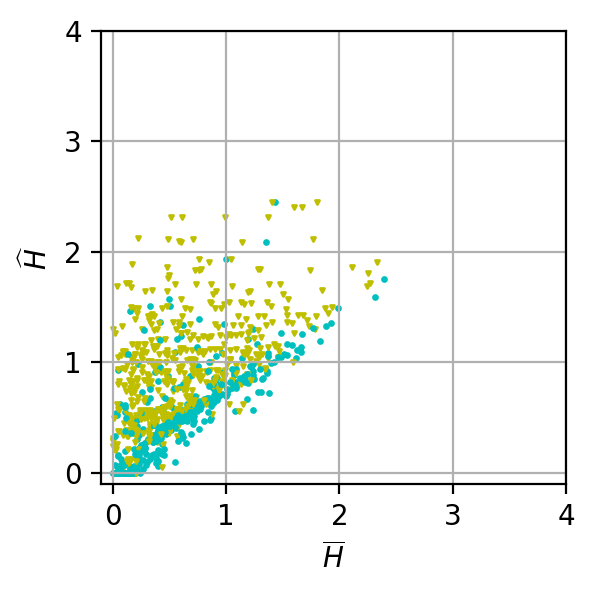}\\
	\end{tabular}
	\caption{Diagrammes de dispersion  $(\overline{\Hv},\hat{\Hv})$  des hauteurs des pics,  parcimonie variable pour $D0$, $D1$ et $D2$ (1re, 2e et 3e lignes),  pour 
		\textbf{U-HQ}, \textbf{U-ISTA} et \textbf{U-PD}  de gauche à droite.}
	\label{fig:comparison_H_sparsity}
\end{figure}

La figure ~\ref{fig:comparison_H_sparsity} détaille l'évolution des hauteurs prédites. La comparaison par colonnes confirme les bonnes performances de \textbf{U-HQ}, et indique que \textbf{U-ISTA} et  \textbf{U-PD} souffrent apparement du biais  $\ell_1$ typique, avec une tendance à sous-estimer une portion importance des hauteurs de pics.

\begin{table*}[hbt]
	\sisetup{round-precision = 2}
	\centering
	\caption{Comparaison des méthodes dépliées \textbf{U-HQ}, \textbf{U-ISTA}, \textbf{U-PD} à $K=4$ couches: moyenne (écart-type) des MSE, SNR et TSNR entre $\pv$ et $\widehat{\pv}$;  moyenne (écart-type) des NMAE pour les hauteurs  $(\overline{\Hv},\widehat{\Hv})$, aires $(\overline{\Av},\widehat{\Av})$ et  localisations $(\overline{\Lv},\widehat{\Lv})$. Meilleures performances en gras.}
	\scalebox{0.81}{
		\begin{tabular}{|c|c|c|c|c|c|c|c|}
			\hline
			& &MSE&SNR&TSNR&$\text{NMAE}(\bar{\Hv},\hat{\Hv})$& $\text{NMAE}(\bar{\Av},\hat{\Av})$& $\text{NMAE}(\bar{\Lv},\hat{\Lv})$\\
			\hline
			\hline
			\parbox[t]{2mm}{\multirow{3}{*}{\rotatebox[origin=c]{90}{\textbf{D0}}}} &\textbf{U-HQ} &\textbf{\num{4.6998681500554085 e-4} (\num{1.1539244587766007e-4})}&\textbf{\num{19.599210739135742} (\num{0.945231556892395})}&\textbf{\num{19.9929} (\num{0.98022})}&\textbf{\num{0.08988} 
				(\num{0.04059})}&\textbf{\num[round-mode=places,round-precision=3]{0.01180} (\num[round-mode=places,round-precision=3]{0.00251})}&\textbf{\num{3.11952 1e-6} (\num{1.45424 1e-6})}\\
			&\textbf{U-PD}&\num{3.684008610932362e-3} (\num{1.6041620968092261e-3})&\num{ 10.882434871919282} (\num{1.3045282031280065})&\num{10.9116} (\num{ 1.3155})&\num{0.2313} (\num{0.04566}) &\num[round-mode=places,round-precision=3]{ 0.10674} (\num[round-mode=places,round-precision=3]{0.024791})&\num{8.04115 1e-6} (\num{1.76754 1e-6})\\
			&\textbf{U-ISTA}&\num{2.988627605839009e-3} (\num{1.0342580416832796e-3})&\num{ 11.681864590364837} (\num{0.9918959207341136})&\num{11.7059} (\num{0.99440})&\num{0.23093} (\num{0.04436})&\num[round-mode=places,round-precision=3]{0.11044} (\num[round-mode=places,round-precision=3]{0.020492})&\num{8.02155 1e-6} (\num{1.70411 1e-6})\\
			\hline 
			\hline
			\parbox[t]{2mm}{\multirow{3}{*}{\rotatebox[origin=c]{90}{\textbf{D1}}}}      &\textbf{ U-HQ} &\textbf{\num{1.0610097087919712e-3}} (\textbf{\num{2.2839375014882535 e-4}})&\textbf{\num{19.356121063232422}} (\textbf{\num{0.6795204877853394}})&\textbf{\num{19.6013}} (\textbf{\num{0.7055}}) &\textbf{\num{0.21749}} (\textbf{\num{0.051295}})&\textbf{\num[round-mode=places,round-precision=3]{0.014193}} (\textbf{\num[round-mode=places,round-precision=3]{0.00219}})&\textbf{\num{3.81178 1e-6}} (\textbf{\num{9.4592 1e-7}})\\
			&\textbf{ U-PD}&\num{ 9.966708622009459e-3} (\num{2.5125569956862574e-3})&\num{9.667551327137815} (\num{0.7117514875628972})&\num{9.6976} (\num{0.7111})&\num{ 0.361302} (\num{0.04284})&\num[round-mode=places,round-precision=3]{0.10677} (\num[round-mode=places,round-precision=3]{0.01596})&\num{6.3290 1e-6} (\num{8.6874 1e-7})\\
			&\textbf{U-ISTA}&\num{8.021826278269716e-3} (\num{2.062176547783416e-3})&\num{10.613744017046706} (\num{0.7296007309453737})&\num{10.6280} (\num{0.7313})&\num{0.34959} (\num{0.0443})&\num[round-mode=places,round-precision=3]{0.10884} (\num[round-mode=places,round-precision=3]{0.0190})&\num{6.12058 1e-6} (\num{8.5869 1e-7})\\
			\hline
			\hline
			\parbox[t]{2mm}{\multirow{3}{*}{\rotatebox[origin=c]{90}{\textbf{D2}}}} &\textbf{ U-HQ} &\textbf{\num{1.6851357650011778 e-3}} (\textbf{\num{3.3682663342915475e-4}})&\textbf{\num{19.55959701538086}} (\textbf{\num{0.7582183480262756}})&\textbf{\num{19.7359}} (\textbf{\num{0.78479}}) &\textbf{\num{0.44525}} (\textbf{\num{0.08008}})&\textbf{\num[round-mode=places,round-precision=3]{0.014767}} (\textbf{\num[round-mode=places,round-precision=3]{0.00194}})&\textbf{\num{5.17277 1e-6}} (\textbf{\num{9.54636 1e-7}}) \\
			&\textbf{ U-PD}&\num{1.721956045744339e-2} (\num{3.2531756286225473e-3})&\num{9.456952338769538} (\num{0.5395920887417428})&\num{9.4921} (\num{0.54033})&\num{0.50243} (\num{ 0.0609})&\num[round-mode=places,round-precision=3]{0.10138} (\num[round-mode=places,round-precision=3]{0.01221 1})&\num{5.8342 1e-6} (\num{7.2260 1e-7})\\
			&\textbf{ U-ISTA}&\num{1.377610915322888e-2} (\num{2.603326007076221e-3})&\num{10.42694260567803} (\num{0.5504398026606626})&\num{10.44122} (\num{0.5504})&\num{0.4970} (\num{0.06304})&\num[round-mode=places,round-precision=3]{0.10249} (\num[round-mode=places,round-precision=3]{0.011775})&\num{5.7721 1e-6} (\num{7.5058 1e-7})\\
			\hline\hline
			\parbox[t]{2mm}{\multirow{3}{*}{\rotatebox[origin=c]{90}{\textbf{D3}}}}&\textbf{ U-HQ} &\textbf{\num{8.73585871886462 e-4}} (\textbf{\num{2.130734792444855e-4}})&\textbf{\num{16.91939353942871}} (\textbf{\num{0.7434666752815247}})&\textbf{\num{17.2977}} (\textbf{\num{0.76735}})&\textbf{\num{0.10442}} (\textbf{\num{0.05026}})&\textbf{\num[round-mode=places,round-precision=3]{0.01466}} (\textbf{\num[round-mode=places,round-precision=3]{0.00229}})&\textbf{\num{3.6768 1e-6}} (\textbf{\num{1.7387 1e-6}})\\
			
			&\textbf{U-PD}&\num{4.350600070162037e-3} (\num{1.366519016468491e-3})&\num{10.03655330531525} (\num{0.8511903413553235})&\num{10.0778} (\num{0.8588})&\num{0.2554} (\num{0.04601})&\num[round-mode=places,round-precision=3]{0.10655} (\num[round-mode=places,round-precision=3]{0.02047})&\num{9.02459 1e-6} (\num{ 1.73923 1e-6})\\
			
			&\textbf{U-ISTA}&\num{3.7013347843635625e-3 } (\num{1.1017852308670168e-3})&\num{10.708997124207022} (\num{0.6549534141213141})&\num{10.7481} (\num{0.66105})&\num{0.24698} (\num{0.04496})&\num[round-mode=places,round-precision=3]{0.10420} (\num[round-mode=places,round-precision=3]{0.02652})&\num{8.7302 1e-6} (\num{1.71667 1e-6})\\
			\hline
			\hline

			\parbox[t]{2mm}{\multirow{3}{*}{\rotatebox[origin=c]{90}{\textbf{D4}}}} &\textbf{U-HQ} &\textbf{\num{1.356011722236871 e-3}} (\textbf{\num{2.8397029382176697 e-4}})&\textbf{\num{ 14.894014358520508}} (\textbf{\num{0.57097989320755}})&\textbf{\num{15.1114}} (\textbf{\num{0.5692}})&\textbf{\num{0.1286}} (\textbf{\num{0.05413}})&\textbf{\num[round-mode=places,round-precision=3]{0.01222}} (\textbf{\num[round-mode=places,round-precision=3]{0.00211}})&\textbf{\num{4.4853 1e-6}} (\textbf{\num{1.88027 1e-6}})\\
			
			&\textbf{U-PD}&\num{5.007966503451674e-3} (\num{1.3578939283909884e-3})&\num{9.284540752222513} (\num{0.509157795677411})&\num{9.3189} (\num{0.51758})&\num{0.27354} (\num{0.04359})&\num{0.09861} (\num{0.01508})&\num{9.5609 1e-6} (\num{1.64512 1e-6})\\
			&\textbf{ U-ISTA}&\num{ 4.483596969892074e-3 } (\num{1.1931322094413733e-3})&\num{9.76272994239367} (\num{0.551096410766514})&\num{9.82679} (\num{0.55110})&\num{0.26513} (\num{0.04414})&\num[round-mode=places,round-precision=3]{0.08326} (\num[round-mode=places,round-precision=3]{0.010749})&\num{ 9.26212 1e-6} (\num{1.635198 1e-6})\\
			
			\hline
			\hline
			
			\parbox[t]{2mm}{\multirow{3}{*}{\rotatebox[origin=c]{90}{\textbf{D5}}}}&\textbf{U-HQ} &\textbf{\num{1.511517446488142 e-3}} (\textbf{\num{3.2763509079813957 e-4}})&\textbf{\num{17.77057647705078}} (\textbf{\num{0.749595582485199}})&\textbf{\num{18.0217}} (\textbf{\num{0.77795}}) &\textbf{\num{0.2379}} (\textbf{\num{0.05653}}) &\textbf{\num[round-mode=places,round-precision=3]{0.01811}} (\textbf{\num[round-mode=places,round-precision=3]{0.002668}})&\textbf{\num{4.18567 1e-6}} (\textbf{\num{1.04700 1e-6}})\\
			
			&\textbf{U-PD}&\num{9.961118058476699 e-3} (\num{2.3353368399516225e-3})&\num{9.60242735665714} (\num{0.6520324069415293})&\num{9.64162} (\num{0.65969})&\num{0.36772} (\num{0.04588})&\num[round-mode=places,round-precision=3]{0.107577} (\num[round-mode=places,round-precision=3]{0.01524})&\num{6.46681 1e-6} (\num{9.09901 1e-7})\\
			
			&\textbf{U-ISTA}&\num{7.880818385762257e-3} (\num{1.7603220079935036e-3})&\num{10.60844900844585} (\num{0.6243645045993629})&\num{10.6240} (\num{0.62532})&\num{0.35831} (\num{ 0.0465})&\num[round-mode=places,round-precision=3]{0.106033} (\num[round-mode=places,round-precision=3]{0.014218})&\num{6.30035 1e-6} (\num{9.10376 1e-7})\\
			
			\hline
			\hline
			
			\parbox[t]{2mm}{\multirow{3}{*}{\rotatebox[origin=c]{90}{\textbf{D6}}}}&\textbf{U-HQ} &\textbf{\num{1.903958385810256e-3}} (\textbf{\num{4.004175425507128e-4}})&\textbf{\num{16.790861129760742}} (\textbf{\num{0.7631433606147766}})&\textbf{\num{17.10528}} (\textbf{\num{0.79613}}) &\textbf{\num{0.2380}} (\textbf{\num{0.05288}})&\textbf{\num[round-mode=places,round-precision=3]{0.021455}} (\textbf{\num[round-mode=places,round-precision=3]{0.003136}})&\textbf{\num{4.17896 1e-6}} (\textbf{\num{9.78106 1e-7}})\\
			
			&\textbf{U-PD}&\num{9.915716886552942e-3} (\num{2.315793044589212e-3})&\num{ 9.651982963402192} (\num{0.6589685580520587})&\num{9.6836} (\num{0.66424})&\num{0.3572} (\num{0.04406})&\num[round-mode=places,round-precision=3]{0.10758} (\num[round-mode=places,round-precision=3]{0.01613})&\num{6.2691 1e-6} (\num{8.69618 1e-7})\\
			
			&\textbf{U-ISTA}&\num{7.851974223145828e-3} (\num{1.7126936316754248e-3})&\num{10.6489968725796} (\num{ 0.6203500561808107})&\num{10.6669} (\num{0.62046})&\num{0.34753} (\num{0.04372})&\num[round-mode=places,round-precision=3]{0.10697} (\num[round-mode=places,round-precision=3]{0.014100})&\num{6.09908 1e-6} (\num{8.59119 1e-7})\\
			\hline 
		\end{tabular}
	}
	\label{tab:arb_layers_MSE_SNR_TSNR}
\end{table*}



\section{Conclusions}
Nous avons proposé une analyse comparative de trois algorithmes de restauration en architecture dépliée : primal-dual, ISTA (seuillage doux itératif) et semi-quadratique. L'application choisie est le traitement  de signaux chromatographiques parcimonieux. Cette analyse se base sur la réalisation d'une boîte à outils permettant de générer différents  ensembles de données chromatographiques paramétrées. Les résultats expérimentaux soulignent l'aspect pratique des architectures déroulées face à des données de complexité graduelle, leur robustesse, leur flexibilité et leur efficacité. Il s'agit d'un domaine prometteur puisqu'il peut permettre aux chimistes de traiter leurs spectres obtenus rapidement et efficacement à l'aide d'outils d'optimisation sophistiqués. Les codes Python sont disponibles\footnote{\href{https://github.com/GHARBIMouna/Unrolled-Half-Quadratic/}{https://github.com/GHARBIMouna/Unrolled-Half-Quadratic/}}.


\end{document}